\documentclass[%
 reprint,
 amsmath,amssymb,
 aps,
]{revtex4-1}

\usepackage{graphicx}% Include figure files
\usepackage{dcolumn}% Align table columns on decimal point
\usepackage{bm}% bold math
\usepackage{longtable}
\usepackage{enumerate}
\usepackage{subfigure}
\usepackage{multirow}
\usepackage{cancel}
\usepackage{tabularx}
\usepackage{threeparttable}
\usepackage{dcolumn}
\usepackage{amssymb}

\newcommand{\Lower}[1]{\smash{\lower 1.5ex \hbox{#1}}}
\begin{document}

\preprint{APS/123-QED}
\title{Efficient description of strongly correlated electrons with mean-field cost}% Force line breaks with \\
%\thanks{A footnote to the article title}%
\author{Katharina Boguslawski}
% \altaffiliation[Also at ]{Physics Department, XYZ University.}%Lines break automatically or can be forced with \\
\author{Pawe{\l} Tecmer}%
\author{Paul W. Ayers}
\email{ayers@mcmaster.ca}
\affiliation{%
 Department of Chemistry and Chemical Biology, McMaster University, Hamilton, 1280 Main Street West, L8S 4M1, Canada \\
% This line break forced with \textbackslash\textbackslash
}%
\vspace{-2cm}
 \author{Patrick Bultinck}
%\author{+ XXX}
\affiliation{%
Department of Inorganic and Physical Chemistry, Ghent University, Krijgslaan 281 (S3), 9000 Gent, Belgium
 \\}
 \vspace{-2cm}
\author{Stijn De Baerdemacker}
\author{Dimitri Van Neck}
\email{dimitri.vanneck@ugent.be}
\affiliation{%
Center for Molecular Modelling, Ghent University, Technologiepark 903, 9052 Gent, Belgium
 \\}
% This line break forced with \textbackslash\textbackslash
%}%
%%\collaboration{MUSO Collaboration}%\noaffiliation
%%
%%\author{Charlie Author}
%% \homepage{http://www.Second.institution.edu/~Charlie.Author}
%%\affiliation{
%% Second institution and/or address\\
%% This line break forced% with \\
%%}%
%%\affiliation{
%% Third institution, the second for Charlie Author
%%}%
%%\author{Delta Author}
%%\affiliation{%
%% Authors' institution and/or address\\
%% This line break forced with \textbackslash\textbackslash
%%}%

%%\collaboration{CLEO Collaboration}%\noaffiliation

\date{\today}% It is always \today, today,
             %  but any date may be explicitly specified

\begin{abstract}
We present an efficient approach to the electron correlation problem that is well-suited for strongly interacting many-body systems, but requires only mean-field-like computational cost. 
The performance of our approach is illustrated for the one-dimensional Hubbard model for different ring lengths, and for the non-relativistic quantum chemical Hamiltonian exploring the symmetric dissociation of the H$_{50}$ hydrogen chain.

%\begin{description}
%\item[Usage]
%Secondary publications and information retrieval purposes.
%\item[PACS numbers]
%May be entered using the \verb+\pacs{#1}+ command.
%\item[Structure]
%You may use the \texttt{description} environment to structure your abstract;
%use the optional argument of the \verb+\item+ command to give the category of each item. 
%\end{description}
\end{abstract}

\pacs{Valid PACS appear here}% PACS, the Physics and Astronomy
                             % Classification Scheme.
%\keywords{Suggested keywords}%Use showkeys class option if keyword
                              %display desired
\maketitle

%\tableofcontents
%%%%%%%%%%%%%%%%%%%%%%%%%%%%%%%%%%%%%%%%%%%%%%%%%%%%%%%%%%%%%%%%%%%%%%%%%%%%%%%%%%%%%%%%%%%%%%%%%%%%%%%%%%%%%%%%%%%%%%%%%%%%%%%%%%%%%%%%%%%%%%%%%%%
%%%%%%%%   INTRODUCTION  %%%%%%%%%%%%%%%%%%%%%%%%%%%%%%%%%%%%%%%%%%%%%%%%%%%%%%%%%%%%%%%%%%%%%%%%%%%%%%%%%%%%%%%%%%%%%%%%%%%%%%%%%%%%%%%%%%%%%%%%%%
%%%%%%%%%%%%%%%%%%%%%%%%%%%%%%%%%%%%%%%%%%%%%%%%%%%%%%%%%%%%%%%%%%%%%%%%%%%%%%%%%%%%%%%%%%%%%%%%%%%%%%%%%%%%%%%%%%%%%%%%%%%%%%%%%%%%%%%%%%%%%%%%%%%
The accurate description of the electron--electron interaction at the quantum-mechanical level is a key problem in condensed matter physics and quantum chemistry.
Since most of the quantum many-body problems are extraordinarily difficult to solve exactly, different approximation schemes emerged \cite{Foulkes2001,monika_mrcc,chanreview,Pollet2012,Hubbard-Gustavo_2013}, among which the density matrix renormalization group (DMRG) algorithm~\cite{white,scholl05,ors_springer} gained a lot of popularity in both condensed matter physics~\cite{scholl05} and quantum chemistry~\cite{marti2010b,fenoDMRG,entanglement_letter,DMRG_polarizability,kurashige2013,entanglement_bonding_2013,CUO_DMRG,CheMPS2} over the last decade. 
Since the DMRG algorithm optimizes a matrix product state wavefunction, it is optimally suited for one-dimensional systems; though DMRG studies on higher-dimensional and compact systems have been reported \cite{Chung2000,fenoDMRG,kurashige2013,CUO_DMRG}. Yet, novel theoretical approaches are desirable that can accurately describe strong correlation effects between electrons where the dimension of the Hilbert space exceeds the present-day limit of DMRG or general tensor-network approaches \cite{TTN_Ors} allowing approximately 100 sites or 60 (spatial) orbitals, respectively.

Another promising approach, suitable for larger strongly-correlated electronic systems, uses geminals (two-electron basis functions) as building blocks for the wavefunction~\cite{Hurley_1953,Coleman_1965,Silver_1969,Ortiz_1981,Surjan_1999,Kutzelnigg2012,Surjan_2012,Ellis_2013}. 
One of the simplest practical geminal approaches is the antisymmetric product of 1-reference-orbital geminals (AP1roG)~\cite{Limacher_2013,Piotrus_PT2,Piotrus_Mol-Phys}. 
Unique among geminal methods, AP1roG can be rewritten as a fully general pair-coupled-cluster doubles wavefunction~\cite{p-CCD}, i.e.
%%%%%%%%%%%%%%%%%%%%%%%%%%%%%%%%%%%%%%%%%%%%%%%%%%%%%%%%%%%%%%%%%%%%%%%%%%%%%%%%%%%%%%%%%%%%%%%%%%%%%%%%%%%%%%%%%%%%%%%%%%%%%%%%%%%%%%%%%%%%
%%%%%%%%%%%%%%%%%%%%%% AP1roG wavefunction %%%%%%%%%%%%%%%%%%%%%%%%%%%%%%%%%%%%%%%%%%%%%%%%%%%%%%%%%%%%%%%%%%%%%%%%%%%%%%%%%%%%%%%%%%%%%%%%%
\begin{equation}\label{eq:ap1rog}
|\Psi_{\rm AP1roG}\rangle = \exp \left (  \sum_{i=1}^P \sum_{a=P+1}^K c_i^a a_{a \uparrow}^{\dagger}  a_{a\downarrow}^{\dagger} a_{i\downarrow}  a_{i\uparrow}\right )|\Phi_0 \rangle,
\end{equation}
%%%%%%%%%%%%%%%%%%%%%%%%%%%%%%%%%%%%%%%%%%%%%%%%%%%%%%%%%%%%%%%%%%%%%%%%%%%%%%%%%%%%%%%%%%%%%%%%%%%%%%%%%%%%%%%%%%%%%%%%%%%%%%%%%%%%%%%%%%%%
%%%%%%%%%%%%%%%%%%%%%%%%%%%%%%%%%%%%%%%%%%%%%%%%%%%%%%%%%%%%%%%%%%%%%%%%%%%%%%%%%%%%%%%%%%%%%%%%%%%%%%%%%%%%%%%%%%%%%%%%%%%%%%%%%%%%%%%%%%%%
where $a_{p\sigma}^{\dagger}$ and $a_{p\sigma}$ ($\sigma=\,\downarrow, \uparrow$) are the fermionic creation and annihilation operators, and $|\Phi_0 \rangle$ is some independent-particle wavefunction (usually the Hartree--Fock determinant). Indices $i$ and $a$ correspond to virtual and occupied sites (orbitals) with respect to $|\Phi_0 \rangle$, $P$ and $K$ denote the number of electron pairs ($P=N/2$ with $N$ being the total number of electrons) and orbitals, respectively, and $\{c_i^a\}$ are the geminal coefficients. 
This wavefunction ansatz is size-extensive and has mean-field scaling, $\mathcal{O}\big(P^2(K-P)^2\big)$ for the projected Schr\"odinger equation approach~\cite{Limacher_2013}.

To ensure size-consistency, however, it is necessary to optimize the one-electron basis functions \cite{Limacher_2013}, where all non-redundant orbital rotations span the occupied--occupied, occupied--virtual, and virtual--virtual blocks with respect to the reference Slater determinant $|\Phi_0 \rangle$.
We have implemented a quadratically convergent algorithm: we minimize the energy with respect to the choice of the one-particle basis functions, subject to the constraint that the projected Schr\"{o}dinger equations for the geminal coefficients hold.
Specifically, we use a Newton--Raphson optimizer and a diagonal approximation of the orbital Hessian to obtain the rotated set of orbital expansion coefficients. 
Our algorithm is analogous to the orbital-optimized coupled cluster approach~\cite{Helgaker_book,Scuseria1987,Bozkaya_2011}.
Due to the four-index transformation of the electron repulsion integrals, the computational scaling deteriorates to $\mathcal{O}\big(K^5\big)$.
The orbital-optimized AP1roG (OO-AP1roG) approach was implemented in a developer version of the \textsc{Horton} program package~\cite{HORTON13}.

%%%%%%%%%%%%%%%%%%%%%%%%%%%%%%%%%%%%%%%%%%%%%%%%%%%%%%%%%%%%%%%%%%%%%%%%%%%%%%%%%%%%%%%%%%%%%%%%%%%%%%%%%%%%%%%%%%%%%%%%%%%%%%%%%%%%%%%%%%%%%%%%%%%
%%%%%%%%   HUBBARD       %%%%%%%%%%%%%%%%%%%%%%%%%%%%%%%%%%%%%%%%%%%%%%%%%%%%%%%%%%%%%%%%%%%%%%%%%%%%%%%%%%%%%%%%%%%%%%%%%%%%%%%%%%%%%%%%%%%%%%%%%%
%%%%%%%%%%%%%%%%%%%%%%%%%%%%%%%%%%%%%%%%%%%%%%%%%%%%%%%%%%%%%%%%%%%%%%%%%%%%%%%%%%%%%%%%%%%%%%%%%%%%%%%%%%%%%%%%%%%%%%%%%%%%%%%%%%%%%%%%%%%%%%%%%%%
\textbf{(a)} \textbf{The half-filled one-dimensional Hubbard Hamiltonian.} 
First, we consider the 1-D Hubbard model Hamiltonian with periodic boundary conditions,
%%%%%%%%%%%%%%%%%%%%%%%%%%%%%%%%%%%%%%%%%%%%%%%%%%%%%%%%%%%%%%%%%%%%%%%%%%%%%%%%%%%%%%%%%%%%%%%%%%%%%%%%%%%%%%%%%%%%%%%%%%%%%%%%%%%%%%%%%%%%
%%%%%%%%%%%%%%%%%%%%%% Hubbard Hamiltonian  %%%%%%%%%%%%%%%%%%%%%%%%%%%%%%%%%%%%%%%%%%%%%%%%%%%%%%%%%%%%%%%%%%%%%%%%%%%%%%%%%%%%%%%%%%%%%%%%
\begin{equation}
\hat{H}_{\rm Hub} = -t\sum_{j,\sigma} \left( a_{(j+1)\sigma}^{\dagger}a_{j\sigma} + a_{j\sigma}^{\dagger}a_{(j+1)\sigma} \right ) 
+U\sum_j n_{j\uparrow} n_{j\downarrow},
\end{equation}
%%%%%%%%%%%%%%%%%%%%%%%%%%%%%%%%%%%%%%%%%%%%%%%%%%%%%%%%%%%%%%%%%%%%%%%%%%%%%%%%%%%%%%%%%%%%%%%%%%%%%%%%%%%%%%%%%%%%%%%%%%%%%%%%%%%%%%%%%%%%
%%%%%%%%%%%%%%%%%%%%%%%%%%%%%%%%%%%%%%%%%%%%%%%%%%%%%%%%%%%%%%%%%%%%%%%%%%%%%%%%%%%%%%%%%%%%%%%%%%%%%%%%%%%%%%%%%%%%%%%%%%%%%%%%%%%%%%%%%%%%
where the first term represents nearest-neighbor hopping and the second term is the repulsive on-site interaction. 
The operators $a_{j\sigma}^{\dagger}$ and $a_{j\sigma}$ are again the fermionic creation and annihilation operators on a lattice with sites $j$ = 1,...,N$_{\rm sites}$, and $n_{j\sigma} = a_{j\sigma}^{\dagger}a_{j\sigma}$ is the local number operator.

%%%%%%%%%%%%%%%%%%%%%%%%%%%%%%%%%%%%%%%%%%%%%%%%%%%%%%%%%%%%%%%%%%%%%%%%%%%%%%%%%%%%%%%%%%%%%%%%%%%%%%%%%%%%%%%%%%%%%%%%%%%%%%%%%%%%%%%%%%%%
%%%%%%%%%%%%%%%%%%% U graph per site energy %%%%%%%%%%%%%%%%%%%%%%%%%%%%%%%%%%%%%%%%%%%%%%%%%%%%%%%%%%%%%%%%%%%%%%%%%%%%%%%%%%%%%%%%%%%%%%%%
\begin{figure}[t]
\centering
\includegraphics[width=0.99\linewidth]{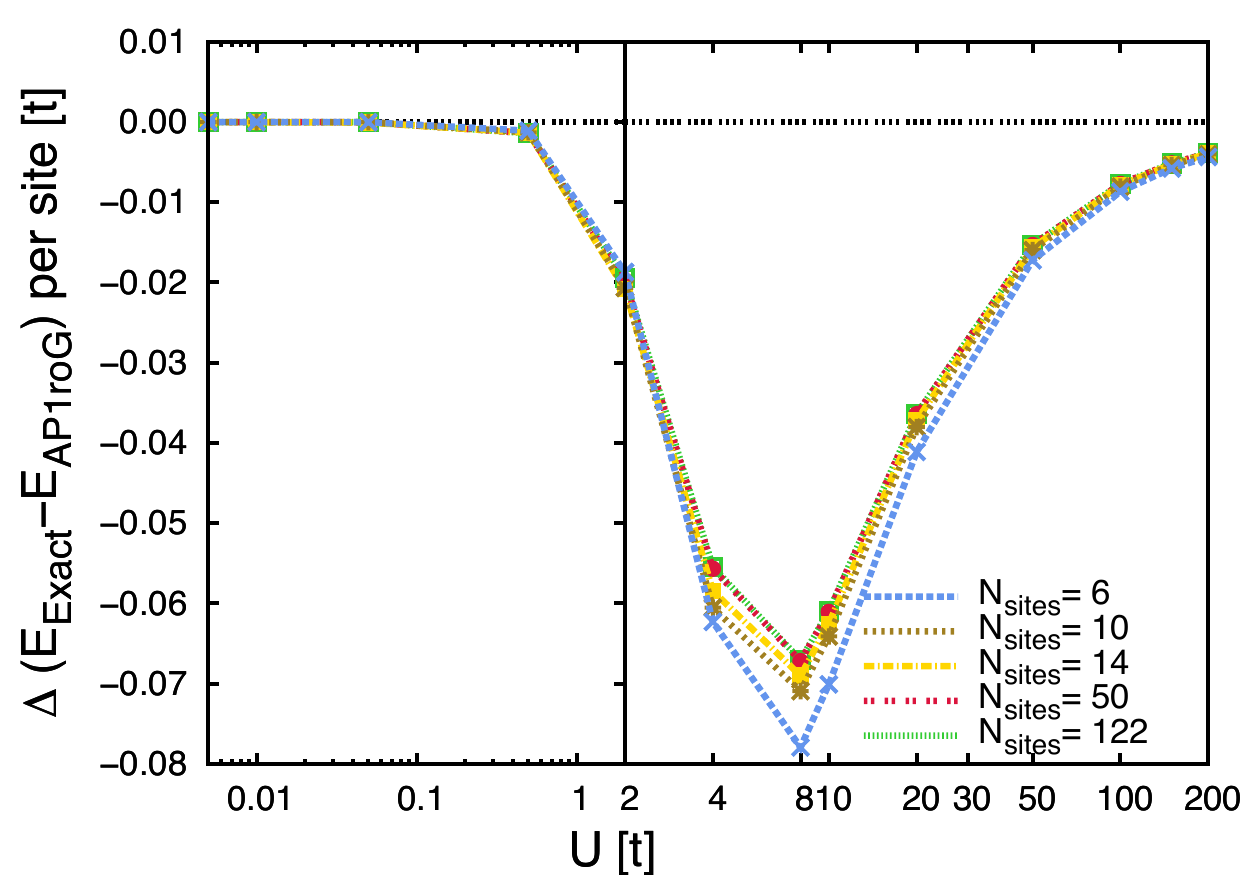}
\caption{Deviation of the OO-AP1roG total energies from exact values (blue dashed line) for different strengths of the repulsive on-site interaction for the 1-D Hubbard model (with periodic boundary conditions) for N$_{\rm sites}=6, 10, 14, 50, 122$. 
The exact values for small $U$ ($U<0.001t$) for N$_{\rm sites}=50, 122$ could not be converged. 
}
\label{fig:diff}
\end{figure}
%%%%%%%%%%%%%%%%%%%%%%%%%%%%%%%%%%%%%%%%%%%%%%%%%%%%%%%%%%%%%%%%%%%%%%%%%%%%%%%%%%%%%%%%%%%%%%%%%%%%%%%%%%%%%%%%%%%%%%%%%%%%%%%%%%%%%%%%%%%%
%%%%%%%%%%%%%%%%%%%%%%%%%%%%%%%%%%%%%%%%%%%%%%%%%%%%%%%%%%%%%%%%%%%%%%%%%%%%%%%%%%%%%%%%%%%%%%%%%%%%%%%%%%%%%%%%%%%%%%%%%%%%%%%%%%%%%%%%%%%%

\begin{figure}[t]
\centering
\includegraphics[width=0.99\linewidth]{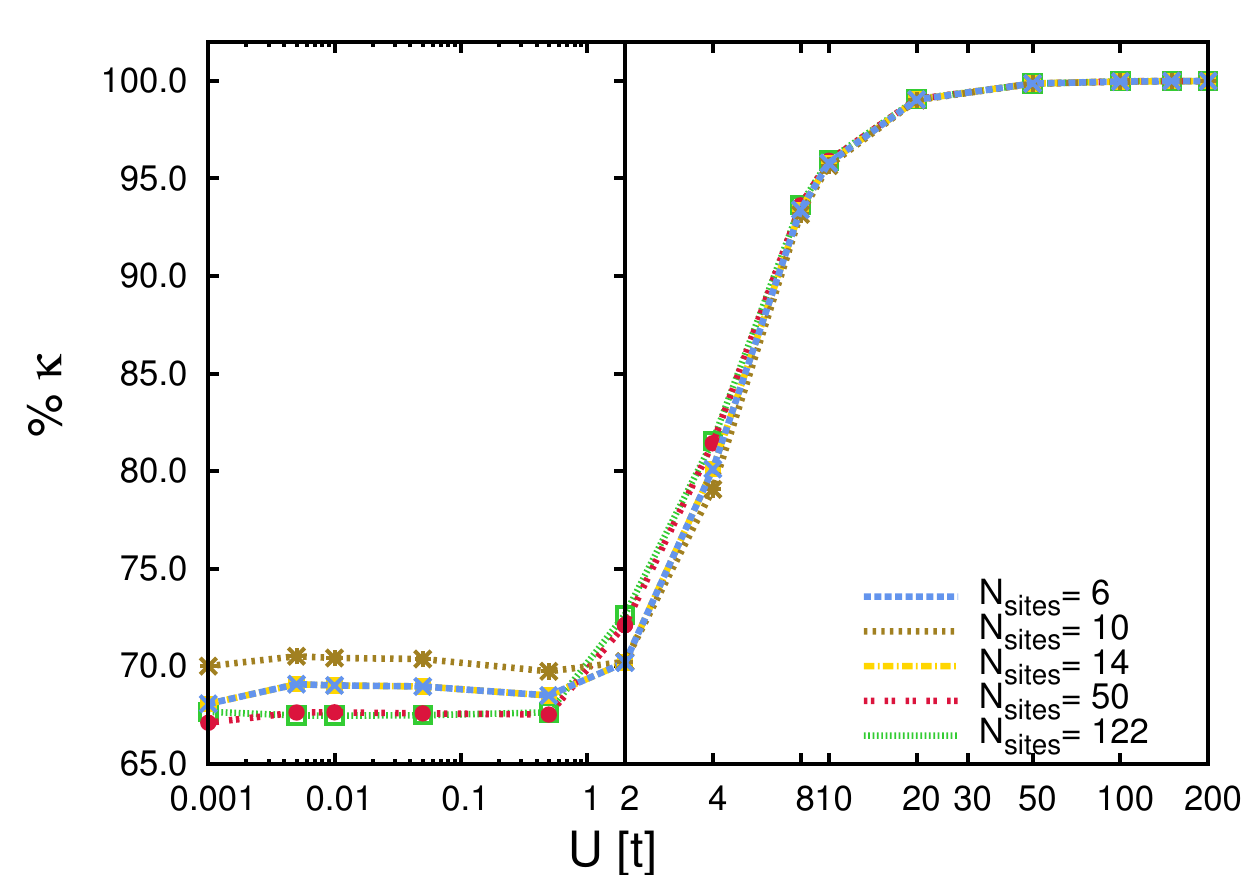}
\caption{Percentage of the correlation energy $\%\kappa$ for different strengths of the repulsive on-site interaction in the half-filled 1-D Hubbard model (with periodic boundary conditions) for N$_{\rm sites}=6, 10, 14, 50, 122$ captured by OO-AP1roG. 
The exact values for small $U$ ($U<0.001t$) for N$_{\rm sites}=50, 122$ could not be converged. 
}
\label{fig:k}
\end{figure}

Figure~\ref{fig:diff} shows the differences in total energies obtained for OO-AP1roG with respect to reference data obtained from the solution of the Lieb-Wu equations~\cite{Lieb-Wu} (N$_{\rm sites}=6,10,14,50,122$). 
OO-AP1roG can reproduce the exact total energies in the limit of zero and infinite (repulsive) on-site interaction.
The largest deviations from the exact solution (up to $0.075t$ per site) are found for the intermediate region of the on-site interaction, that is, for $2t < U < 50t$. Figure~\ref{fig:k} shows the percentage of the correlation energy captured by OO-AP1roG calculated as $\%\kappa = \frac{E^{\rm OO\text{-}AP1roG}-E^{\rm HF}}{E^{\rm exact}-E^{\rm HF}}\cdot 100$. In the limit of zero and infinite $U$, the OO-AP1roG model becomes exact; for $U=0$ the wavefunction can be exactly described by a single Slater determinant and thus the correlation energy approaches zero, while for $U\rightarrow \infty$, the quantum state can be represented by the perfect pairing wavefunction. For growing (repulsive) $U$, the percentage of the correlation energy covered by OO-AP1roG increases gradually.

For small values of $U$, the geminal coefficient matrix $\{c_i^a\}$ is sparse and thus far from perfect pairing, which is represented by a diagonal geminal coefficient matrix (see Figure~I in the Supplementary Information). In the limit $U\rightarrow 0$, the geminal coefficient matrix correctly approaches the zero matrix indicating that a single Slater determinant is sufficient to describe the quantum state exactly. For increasing $U$, $\{c_i^a\}$ becomes diagonal-dominant and adopts a diagonal structure in the limit of $U\rightarrow \infty$. Thus, in the limit of infinite (repulsive) interaction, OO-AP1roG optimizes a perfect-pairing (seniority-zero) wavefunction~\cite{Neuscamman_2012,Gus_seniority}, 
%%%%%%%%%%%%%%%%%%%%%%%%%%%%%%%%%%%%%%%%%%%%%%%%%%%%%%%%%%%%%%%%%%%%%%%%%%%%%%%%%%%%%%%%%%%%%%%%%%%%%%%%%%%%%%%%%%%%%%%%%%%%%%%%%%%%%%%%%%%%
%%%%%%%%%%%%%%%%%%%%%%% perfect pairing wavefunction %%%%%%%%%%%%%%%%%%%%%%%%%%%%%%%%%%%%%%%%%%%%%%%%%%%%%%%%%%%%%%%%%%%%%%%%%%%%%%%%%%%%%%
\begin{equation}
\prod_{i=1,3,..}[ (a^\dagger_{i,\uparrow}+a^\dagger_{i+1, \uparrow})(a^\dagger_{i,\downarrow}+a^\dagger_{i+1, \downarrow})
                -(a^\dagger_{i,\uparrow}-a^\dagger_{i+1, \uparrow})(a^\dagger_{i,\downarrow}-a^\dagger_{i+1, \downarrow}) ] |0\rangle
\end{equation}
%%%%%%%%%%%%%%%%%%%%%%%%%%%%%%%%%%%%%%%%%%%%%%%%%%%%%%%%%%%%%%%%%%%%%%%%%%%%%%%%%%%%%%%%%%%%%%%%%%%%%%%%%%%%%%%%%%%%%%%%%%%%%%%%%%%%%%%%%%%%
%%%%%%%%%%%%%%%%%%%%%%%%%%%%%%%%%%%%%%%%%%%%%%%%%%%%%%%%%%%%%%%%%%%%%%%%%%%%%%%%%%%%%%%%%%%%%%%%%%%%%%%%%%%%%%%%%%%%%%%%%%%%%%%%%%%%%%%%%%%%

%%%%%%%%%%%%%%%%%%%%%%%%%%%%%%%%%%%%%%%%%%%%%%%%%%%%%%%%%%%%%%%%%%%%%%%%%%%%%%%%%%%%%%%%%%%%%%%%%%%%%%%%%%%%%%%%%%%%%%%%%%%%%%%%%%%%%%%%%%%%
%%%%%%%%%%%%%%%%%%%%% correlation energy plot  %%%%%%%%%%%%%%%%%%%%%%%%%%%%%%%%%%%%%%%%%%%%%%%%%%%%%%%%%%%%%%%%%%%%%%%%%%%%%%%%%%%%%%%%%%%%%
%%%%%%%%%%%%%%%%%%%%%%%%%%%%%%%%%%%%%%%%%%%%%%%%%%%%%%%%%%%%%%%%%%%%%%%%%%%%%%%%%%%%%%%%%%%%%%%%%%%%%%%%%%%%%%%%%%%%%%%%%%%%%%%%%%%%%%%%%%%%
%%%%%%%%%%%%%%%%%%%%%%%%%%%%%%%%%%%%%%%%%%%%%%%%%%%%%%%%%%%%%%%%%%%%%%%%%%%%%%%%%%%%%%%%%%%%%%%%%%%%%%%%%%%%%%%%%%%%%%%%%%%%%%%%%%%%%%%%%%%%

%Table~\ref{tbl:Hubbard} lists the ground-state (E) and correlation ($\kappa$) energies of half-filled 1-D lattices of different lengths obtained from OO-AP1roG. 
%For each chain length, three different values of $U$, $U=2t,4t$ and $8t$, representing weak, intermediate and strong electron-electron interaction, respectively, have been chosen and compared with the corresponding exact solutions~\cite{Bethe_ansatz, Lieb_1968}. 
To conclude, OO-AP1roG has mean-field-like scaling, but can recover about 71$\%$ of the correlation energy in the weak interaction regime, about 80$\%$ for intermediate interaction strengths, and approximately 93$\%$ in the case of strong on-site interaction for all chain lengths studied (a numerical comparison is presented in Table~I of the Supplementary Information). 

%%%%%%%%%%%%%%%%%%%%%%%%%%%%%%%%%%%%%%%%%%%%%%%%%%%%%%%%%%%%%%%%%%%%%%%%%%%%%%%%%%%%%%%%%%%%%%%%%%%%%%%%%%%%%%%%%%%%%%%%%%%%%%%%%%%%%%%%%%%%
%%%%%%%%%%%%%%%%%% pair entanglement %%%%%%%%%%%%%%%%%%%%%%%%%%%%%%%%%%%%%%%%%%%%%%%%%%%%%%%%%%%%%%%%%%%%%%%%%%%%%%%%%%%%%%%%%%%%%%%%%%%%%%%
\begin{figure}[h]
\centering
\includegraphics[width=0.99\linewidth]{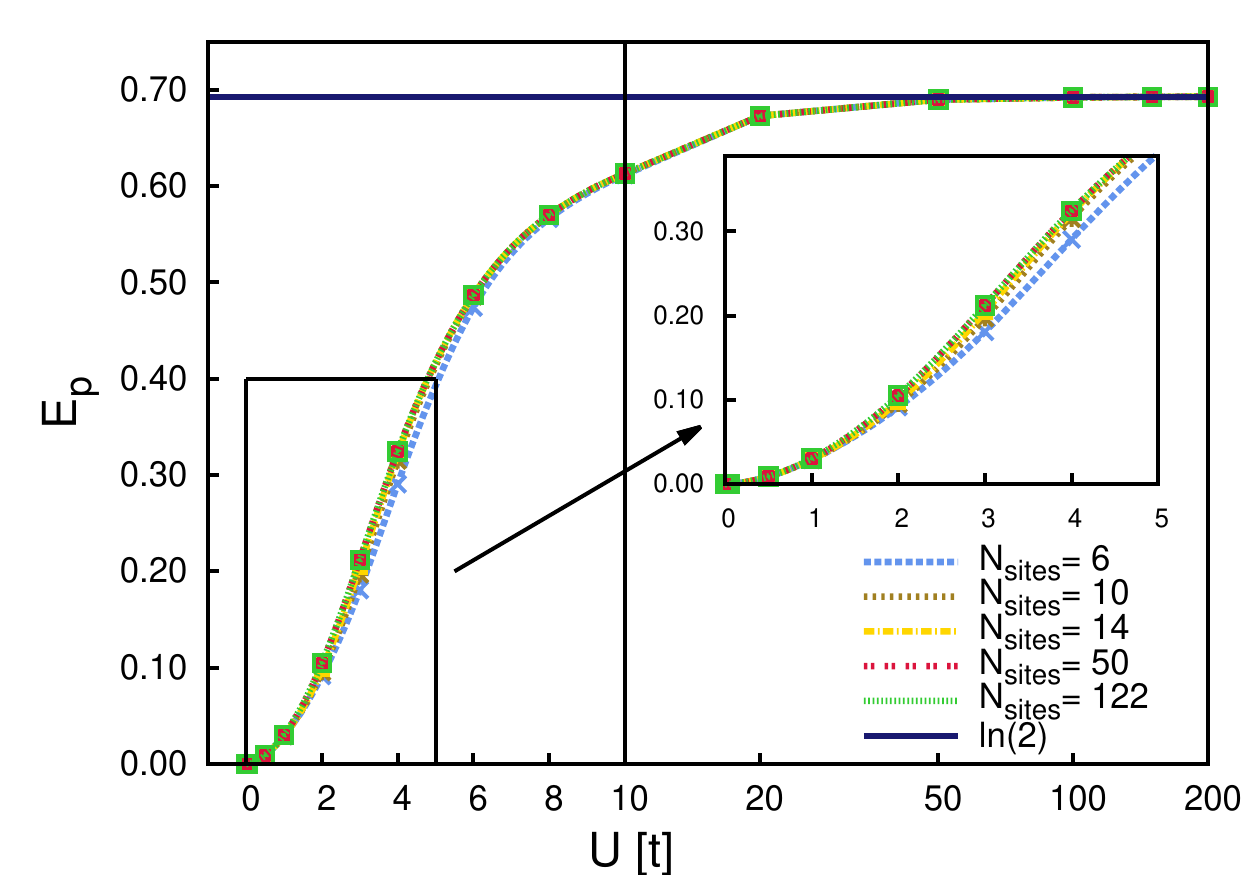}
\caption{Pair entanglement $E_p$ (single-orbital entropy) for different strengths of the repulsive on-site interaction in the half-filled 1-D Hubbard model (with periodic boundary conditions) for N$_{\rm sites}=6, 10, 14, 50, 122$ calculated by OO-AP1roG.
}
\label{fig:site-e}
\end{figure}
%%%%%%%%%%%%%%%%%%%%%%%%%%%%%%%%%%%%%%%%%%%%%%%%%%%%%%%%%%%%%%%%%%%%%%%%%%%%%%%%%%%%%%%%%%%%%%%%%%%%%%%%%%%%%%%%%%%%%%%%%%%%%%%%%%%%%%%%%%%%
%%%%%%%%%%%%%%%%%%%%%%%%%%%%%%%%%%%%%%%%%%%%%%%%%%%%%%%%%%%%%%%%%%%%%%%%%%%%%%%%%%%%%%%%%%%%%%%%%%%%%%%%%%%%%%%%%%%%%%%%%%%%%%%%%%%%%%%%%%%%
Figure~\ref{fig:site-e} shows the single-orbital entropy for different lengths of the 1-D lattice as a function of the repulsive on-site interaction $U$. The single-orbital entropy is the analogue of the one-site entropy, but determined in the natural orbital basis: it is calculated as the von Neumann entropy from a single-orbital density matrix (a many-particle reduced density matrix of one orbital). It measures the entanglement of one orbital with the remaining (N$_{\rm sites}-1$) ones \cite{entanglement_bonding_2013}. In particular, since the optimized orbitals are localized on two neighboring sites, the von Neumann entropy describes the correlation of a pair of sites and the other part of the system. In the following, we will refer to the single-orbital entropy as the pair entanglement $E_p$ in accordance with the local entanglement determined for the on-site basis \cite{Gu2004}.

The pair entanglement takes its minimum value at $U=0t$ where the wavefunction can be exactly represented by a single Slater determinant. It is easy to verify that all orbital pairs are uncorrelated in a one-determinant wavefunction and thus $E_p=0$. For increasing on-site interaction, the pair entanglement smoothly accumulates (see Figure~\ref{fig:site-e}) and reaches its maximum value of $\ln 2$ in the large $U$ limit (for $U\rightarrow \infty$, the single-orbital density matrix has the diagonal elements $\{0,0,0.5,0.5\}$). Note that OO-AP1roG yields similar pair entanglement profiles for all chain lengths studied and correctly reproduces the small and large $U$ limits.

%%%%%%%%%%%%%%%%%%%%%%%%%%%%%%%%%%%%%%%%%%%%%%%%%%%%%%%%%%%%%%%%%%%%%%%%%%%%%%%%%%%%%%%%%%%%%%%%%%%%%%%%%%%%%%%%%%%%%%%%%%%%%%%%%%%%%%%%%%%%%%%%%%%
%%%%%%%%   H50 chain     %%%%%%%%%%%%%%%%%%%%%%%%%%%%%%%%%%%%%%%%%%%%%%%%%%%%%%%%%%%%%%%%%%%%%%%%%%%%%%%%%%%%%%%%%%%%%%%%%%%%%%%%%%%%%%%%%%%%%%%%%%
%%%%%%%%%%%%%%%%%%%%%%%%%%%%%%%%%%%%%%%%%%%%%%%%%%%%%%%%%%%%%%%%%%%%%%%%%%%%%%%%%%%%%%%%%%%%%%%%%%%%%%%%%%%%%%%%%%%%%%%%%%%%%%%%%%%%%%%%%%%%%%%%%%%
\textbf{(b)} \textbf{Symmetric dissociation of the H$_{50}$ molecule.} The non-relativistic quantum chemical Hamiltonian in its second quantized form reads
%%%%%%%%%%%%%%%%%%%%%%%%%%%%%%%%%%%%%%%%%%%%%%%%%%%%%%%%%%%%%%%%%%%%%%%%%%%%%%%%%%%%%%%%%%%%%%%%%%%%%%%%%%%%%%%%%%%%%%%%%%%%%%%%%%%%%%%%%%%%
%%%%%%%%%%%%%%%% chemical Hamiltonian %%%%%%%%%%%%%%%%%%%%%%%%%%%%%%%%%%%%%%%%%%%%%%%%%%%%%%%%%%%%%%%%%%%%%%%%%%%%%%%%%%%%%%%%%%%%%%%%%%%%%%
\begin{equation}\label{eqn:Chem}
\hat{H} = \sum_{pq,\sigma} h_{pq}  a^\dagger_{p\sigma} a_{q\sigma} + \frac{1}{2} \sum_{pqrs,\sigma\tau} \langle pq|rs \rangle a^\dagger_{p\sigma} a^\dagger_{q\tau} a_{s\tau} a_{r\sigma}+H_{\rm nuc},
\end{equation}
%%%%%%%%%%%%%%%%%%%%%%%%%%%%%%%%%%%%%%%%%%%%%%%%%%%%%%%%%%%%%%%%%%%%%%%%%%%%%%%%%%%%%%%%%%%%%%%%%%%%%%%%%%%%%%%%%%%%%%%%%%%%%%%%%%%%%%%%%%%%
%%%%%%%%%%%%%%%%%%%%%%%%%%%%%%%%%%%%%%%%%%%%%%%%%%%%%%%%%%%%%%%%%%%%%%%%%%%%%%%%%%%%%%%%%%%%%%%%%%%%%%%%%%%%%%%%%%%%%%%%%%%%%%%%%%%%%%%%%%%%
where the first term comprises the kinetic energy and nuclear--electron attraction, the second term is the electron-electron interaction, and the third term represents the nuclear--nuclear repulsion energy, respectively.  
In Eq.~\eqref{eqn:Chem}, indices $p$, $q$, $r$ and $s$ run over all one-particle basis functions, while $\sigma$ and $\tau$ denote the electron spin ($\{\uparrow,\downarrow\}$). 
The Hamiltonian as defined in Eq.~\eqref{eqn:Chem} was used for the study of the symmetric stretching of the H$_{50}$ hydrogen chain, which is a commonly-used molecular model for strongly-correlated systems and which remains a challenging problem for conventional quantum-chemistry methods~\cite{Hachmann_H50,Gustavo_H50,Stella_H50,DMFT_H50}.

%%%%%%%%%%%%%%%%%%%%%%%%%%%%%%%%%%%%%%%%%%%%%%%%%%%%%%%%%%%%%%%%%%%%%%%%%%%%%%%%%%%%%%%%%%%%%%%%%%%%%%%%%%%%%%%%%%%%%%%%%%%%%%%%%%%%%%%%%%%%
%%%%%%%%%%%%% H50 dissociation %%%%%%%%%%%%%%%%%%%%%%%%%%%%%%%%%%%%%%%%%%%%%%%%%%%%%%%%%%%%%%%%%%%%%%%%%%%%%%%%%%%%%%%%%%%%%%%%%%%%%%%%%%%%%
\begin{figure}[h]
\centering
\includegraphics[width=0.90\linewidth]{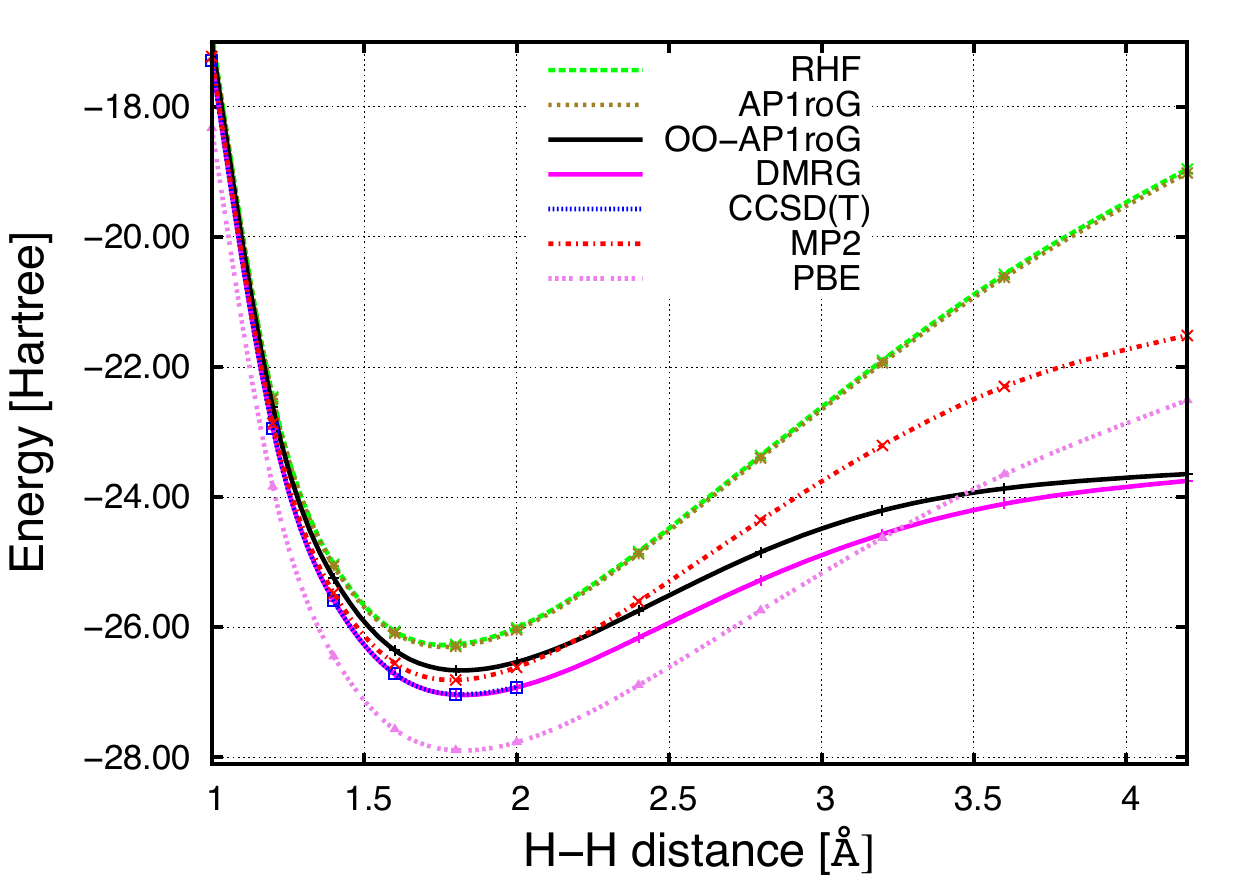}
\caption{Symmetric dissociation of H$_{50}$ chain using the STO-6G basis set~\cite{STO-6G-H-Ne} obtained from different methods. The DMRG reference data are taken from Ref.~\citenum{Hachmann_H50}.
}
\label{fig:H50}
\end{figure}
%%%%%%%%%%%%%%%%%%%%%%%%%%%%%%%%%%%%%%%%%%%%%%%%%%%%%%%%%%%%%%%%%%%%%%%%%%%%%%%%%%%%%%%%%%%%%%%%%%%%%%%%%%%%%%%%%%%%%%%%%%%%%%%%%%%%%%%%%%%%
%%%%%%%%%%%%%%%%%%%%%%%%%%%%%%%%%%%%%%%%%%%%%%%%%%%%%%%%%%%%%%%%%%%%%%%%%%%%%%%%%%%%%%%%%%%%%%%%%%%%%%%%%%%%%%%%%%%%%%%%%%%%%%%%%%%%%%%%%%%%

In Figure~\ref{fig:H50}, the performance of AP1roG and OO-AP1roG is compared to restricted Hartree--Fock (RHF), second-order M{\o}ller-Plesset (MP2) perturbation theory, coupled cluster theory with singles, doubles and perturbative triples (CCSD(T)), and density functional theory using the PBE~\cite{PBEx} exchange--correlation functional. 
As reference, the DMRG potential energy curve determined in Ref.~\citenum{Hachmann_H50} was used, which can be considered as the exact-diagonalization limit. 
None of the standard quantum chemical methods, like MP2, CCSD(T) or DFT using the PBE exchange--correlation functional, yield qualitatively correct energy curves for the symmetric stretching of the H$_{50}$ chain. 
In particular, the potential energy depth determined from DFT and MP2 is too deep, while CCSD(T) does not converge for interatomic distances larger than 2.0 \AA.
Note that the lack of size-consistency in AP1roG is cured by orbital optimization in the OO-AP1roG approach. 
The latter yields a potential energy curve that is closest to the DMRG reference data along the whole dissociation pathway and leads to a proper dissociation limit of H$_{50}$. 
Moreover, the OO-AP1roG method gives spectroscopic constants (presented in Table~\ref{tbl:constants}) that are in excellent agreement with DMRG reference data, outperforming standard quantum-chemistry approaches.

%%%%%%%%%%%%%%%%%%%%%%%%%%%%%%%%%%%%%%%%%%%%%%%%%%%%%%%%%%%%%%%%%%%%%%%%%%%%%%%%%%%%%%%%%%%%%%%%%%%%%%%%%%%%%%%%%%%%%%%%%%%%%%%%%%%%%%%%%%%%
%%%%%%%%%%%%%%% Spectroscopic constants %%%%%%%%%%%%%%%%%%%%%%%%%%%%%%%%%%%%%%%%%%%%%%%%%%%%%%%%%%%%%%%%%%%%%%%%%%%%%%%%%%%%%%%%%%%%%%%%%%%%
\begin{table}[t]
\caption{\small Spectroscopic constants: equilibrium bond distance (R$_{\rm e}$), potential energy depth (D$_{\rm e}$) and harmonic vibrational frequency ($\omega_{\rm e}$) for the ground state of H$_{50}$ (STO-6G). Differences with respect to the DMRG reference data are listed in parenthesis.
}
\label{tbl:constants}
{\scriptsize
\begin{tabular}{llllll} 
%\begin{ruledtabular}{lccc}
\hline \hline
Method &
R$_{\rm e}$ [\AA]&&
D$_{\rm e}$ [eV] &&
$\omega_{\rm e}$ [cm$^{-1}$]\\
\hline
RHF                         &   0.940 ($-$0.030) && 199.0 ($+$109.3) && 25089  ($+$2268)\\
AP1roG                      &   0.941 ($-$0.029) && 198.2 ($+$108.5) && 23013  ($+$2252)\\
MP2                         &   0.955 ($-$0.015) && 144.1 ($+$54.4)  && 24568  ($+$1747)\\
PBE                         &   0.971 ($+$0.001) && 146.6 ($+$56.9)  && 23662  ($+$841) \\
OO-AP1roG                   &   0.966 ($-$0.004) && 82.2  ($-$7.5)   && 23013  ($+$192) \\
DMRG~\cite{Hachmann_H50}    &   0.970            && 89.7             && 22821           \\
\hline \hline
\end{tabular}
}
\end{table}
%%%%%%%%%%%%%%%%%%%%%%%%%%%%%%%%%%%%%%%%%%%%%%%%%%%%%%%%%%%%%%%%%%%%%%%%%%%%%%%%%%%%%%%%%%%%%%%%%%%%%%%%%%%%%%%%%%%%%%%%%%%%%%%%%%%%%%%%%%%%
%%%%%%%%%%%%%%%%%%%%%%%%%%%%%%%%%%%%%%%%%%%%%%%%%%%%%%%%%%%%%%%%%%%%%%%%%%%%%%%%%%%%%%%%%%%%%%%%%%%%%%%%%%%%%%%%%%%%%%%%%%%%%%%%%%%%%%%%%%%%

Wavefunctions constructed as antisymmetric products of nonorthogonal geminals, like the AP1roG wavefunction scrutinized here, provide an alternative approach to electronic structure, with mean-field scaling. Because it uses  electron pairs as a building block, AP1roG is a suitable way to describe strong correlations dominated by electron pairing. 
However, in order to ensure size-consistency, the single-particle (orbital) basis used to construct the electron pairs must be optimized.  
Our results show that orbital-optimized AP1roG is a robust method all the way from the weakly-correlated to the strongly-correlated limit, in both molecules and periodic systems.

%%%%%%%%%%%%%%%%%%%%%%%%%%%%%%%%%%%%%%%%%%%%%%%%%%%%%%%%%%%%%%%%%%%%%%%%%%%%%%%%%%%%%%%%%%%%%%%%%%%%%%%%%%%%%%%%%%%%%%%%%%%%%%%%%%%%%%%%%%%%%%%%%%%
%%%%%%%% ACKNOWLEDGMENTS %%%%%%%%%%%%%%%%%%%%%%%%%%%%%%%%%%%%%%%%%%%%%%%%%%%%%%%%%%%%%%%%%%%%%%%%%%%%%%%%%%%%%%%%%%%%%%%%%%%%%%%%%%%%%%%%%%%%%%%%%%
%%%%%%%%%%%%%%%%%%%%%%%%%%%%%%%%%%%%%%%%%%%%%%%%%%%%%%%%%%%%%%%%%%%%%%%%%%%%%%%%%%%%%%%%%%%%%%%%%%%%%%%%%%%%%%%%%%%%%%%%%%%%%%%%%%%%%%%%%%%%%%%%%%%
\textbf{Acknowledgments}. K.B. acknowledges the financial support from the Swiss National Science Foundation (P2EZP2 148650). P.T. and P.W.A gratefully acknowledge financial support from the Natural Sciences and Engineering Research Council of Canada.
P.B. with S.D.B. and D.V.N acknowledge financial support from FWO-Flanders and the Research Council of Ghent University. S.D.B is an FWO postdoctoral fellow.
We thank Ward Poelmans, Brecht Verstichel and Paul Johnson for results for the Hubbard model obtained by solving the Lieb-Wu equations.
We had many helpful discussions on geminals and exactly-solvable models with Peter Limacher and Paul Johnson.
B.V., H.V.A., W.P., S.W. and D.V.N. are Members of the QCMM alliance Ghent-Brussels.

%%%%%%%%%%%%%%%%%%%%%%%%%%%%%%%%%%%%%%%%%%%%%%%%%%%%%%%%%%%%%%%%%%%%%%%%%%%%%%%%%%%%%%%%%%%%%%%%%%%%%%%%%%%%%%%%%%%%%%%%%%%%%%%%%%%%%%%%%%%%%%%%%%%
%%%%%%%% REFERENCES      %%%%%%%%%%%%%%%%%%%%%%%%%%%%%%%%%%%%%%%%%%%%%%%%%%%%%%%%%%%%%%%%%%%%%%%%%%%%%%%%%%%%%%%%%%%%%%%%%%%%%%%%%%%%%%%%%%%%%%%%%%
%%%%%%%%%%%%%%%%%%%%%%%%%%%%%%%%%%%%%%%%%%%%%%%%%%%%%%%%%%%%%%%%%%%%%%%%%%%%%%%%%%%%%%%%%%%%%%%%%%%%%%%%%%%%%%%%%%%%%%%%%%%%%%%%%%%%%%%%%%%%%%%%%%%
\bibliography{rsc} %your .bib file

%merlin.mbs apsrev4-1.bst 2010-07-25 4.21a (PWD, AO, DPC) hacked
%Control: key (0)
%Control: author (8) initials jnrlst
%Control: editor formatted (1) identically to author
%Control: production of article title (-1) disabled
%Control: page (0) single
%Control: year (1) truncated
%Control: production of eprint (0) enabled
\begin{thebibliography}{50}%
\makeatletter
\providecommand \@ifxundefined [1]{%
 \@ifx{#1\undefined}
}%
\providecommand \@ifnum [1]{%
 \ifnum #1\expandafter \@firstoftwo
 \else \expandafter \@secondoftwo
 \fi
}%
\providecommand \@ifx [1]{%
 \ifx #1\expandafter \@firstoftwo
 \else \expandafter \@secondoftwo
 \fi
}%
\providecommand \natexlab [1]{#1}%
\providecommand \enquote  [1]{``#1''}%
\providecommand \bibnamefont  [1]{#1}%
\providecommand \bibfnamefont [1]{#1}%
\providecommand \citenamefont [1]{#1}%
\providecommand \href@noop [0]{\@secondoftwo}%
\providecommand \href [0]{\begingroup \@sanitize@url \@href}%
\providecommand \@href[1]{\@@startlink{#1}\@@href}%
\providecommand \@@href[1]{\endgroup#1\@@endlink}%
\providecommand \@sanitize@url [0]{\catcode `\\12\catcode `\$12\catcode
  `\&12\catcode `\#12\catcode `\^12\catcode `\_12\catcode `\%12\relax}%
\providecommand \@@startlink[1]{}%
\providecommand \@@endlink[0]{}%
\providecommand \url  [0]{\begingroup\@sanitize@url \@url }%
\providecommand \@url [1]{\endgroup\@href {#1}{\urlprefix }}%
\providecommand \urlprefix  [0]{URL }%
\providecommand \Eprint [0]{\href }%
\providecommand \doibase [0]{http://dx.doi.org/}%
\providecommand \selectlanguage [0]{\@gobble}%
\providecommand \bibinfo  [0]{\@secondoftwo}%
\providecommand \bibfield  [0]{\@secondoftwo}%
\providecommand \translation [1]{[#1]}%
\providecommand \BibitemOpen [0]{}%
\providecommand \bibitemStop [0]{}%
\providecommand \bibitemNoStop [0]{.\EOS\space}%
\providecommand \EOS [0]{\spacefactor3000\relax}%
\providecommand \BibitemShut  [1]{\csname bibitem#1\endcsname}%
\let\auto@bib@innerbib\@empty
%</preamble>
\bibitem [{\citenamefont {Dagotto}(1994)}]{Dagotto_1994}%
  \BibitemOpen
  \bibfield  {author} {\bibinfo {author} {\bibfnamefont {E.}~\bibnamefont
  {Dagotto}},\ }\href@noop {} {\bibfield  {journal} {\bibinfo  {journal} {Rev.
  Mod. Phys.}\ }\textbf {\bibinfo {volume} {66}},\ \bibinfo {pages} {763}
  (\bibinfo {year} {1994})}\BibitemShut {NoStop}%
\bibitem [{\citenamefont {Capelle}\ and\ \citenamefont
  {Campo}(2013)}]{Capelle_rev}%
  \BibitemOpen
  \bibfield  {author} {\bibinfo {author} {\bibfnamefont {K.}~\bibnamefont
  {Capelle}}\ and\ \bibinfo {author} {\bibfnamefont {V.~L.}\ \bibnamefont
  {Campo}},\ }\href@noop {} {\bibfield  {journal} {\bibinfo  {journal} {Phys.
  Report.}\ }\textbf {\bibinfo {volume} {528}},\ \bibinfo {pages} {91}
  (\bibinfo {year} {2013})}\BibitemShut {NoStop}%
\bibitem [{\citenamefont {L\"owdin}(1958)}]{Lowdin_rev}%
  \BibitemOpen
  \bibfield  {author} {\bibinfo {author} {\bibfnamefont {P.-O.}\ \bibnamefont
  {L\"owdin}},\ }in\ \href@noop {} {\emph {\bibinfo {booktitle} {Adv. Chem.
  Phys.}}},\ Vol.~\bibinfo {volume} {I}\ (\bibinfo  {publisher} {Wiley \& Sons,
  Inc},\ \bibinfo {year} {1958})\ Chap.\ \bibinfo {chapter} {Review of
  different approaches and discussion of some current ideas}, pp.\ \bibinfo
  {pages} {209--321}\BibitemShut {NoStop}%
\bibitem [{\citenamefont {Bartlett}\ and\ \citenamefont
  {Musia{\l}}(2007)}]{bartlett_2007}%
  \BibitemOpen
  \bibfield  {author} {\bibinfo {author} {\bibfnamefont {R.~J.}\ \bibnamefont
  {Bartlett}}\ and\ \bibinfo {author} {\bibfnamefont {M.}~\bibnamefont
  {Musia{\l}}},\ }\href@noop {} {\bibfield  {journal} {\bibinfo  {journal}
  {Rev.~Mod.~Phys.}\ }\textbf {\bibinfo {volume} {79}},\ \bibinfo {pages} {291}
  (\bibinfo {year} {2007})}\BibitemShut {NoStop}%
\bibitem [{\citenamefont {Foulkes}\ \emph {et~al.}(2001)\citenamefont
  {Foulkes}, \citenamefont {Mitas}, \citenamefont {Needs},\ and\ \citenamefont
  {Rajagopal}}]{Foulkes2001}%
  \BibitemOpen
  \bibfield  {author} {\bibinfo {author} {\bibfnamefont {W.}~\bibnamefont
  {Foulkes}}, \bibinfo {author} {\bibfnamefont {L.}~\bibnamefont {Mitas}},
  \bibinfo {author} {\bibfnamefont {R.}~\bibnamefont {Needs}}, \ and\ \bibinfo
  {author} {\bibfnamefont {G.}~\bibnamefont {Rajagopal}},\ }\href@noop {}
  {\bibfield  {journal} {\bibinfo  {journal} {Rev. Mod. Phys.}\ }\textbf
  {\bibinfo {volume} {73}},\ \bibinfo {pages} {33} (\bibinfo {year}
  {2001})}\BibitemShut {NoStop}%
\bibitem [{\citenamefont {Lyakh}\ \emph {et~al.}(2012)\citenamefont {Lyakh},
  \citenamefont {Musia{\l}}, \citenamefont {Lotrich},\ and\ \citenamefont
  {Bartlett}}]{monika_mrcc}%
  \BibitemOpen
  \bibfield  {author} {\bibinfo {author} {\bibfnamefont {D.~I.}\ \bibnamefont
  {Lyakh}}, \bibinfo {author} {\bibfnamefont {M.}~\bibnamefont {Musia{\l}}},
  \bibinfo {author} {\bibfnamefont {V.~F.}\ \bibnamefont {Lotrich}}, \ and\
  \bibinfo {author} {\bibfnamefont {.~J.}\ \bibnamefont {Bartlett}},\
  }\href@noop {} {\bibfield  {journal} {\bibinfo  {journal} {Chem.~Rev.}\
  }\textbf {\bibinfo {volume} {112}},\ \bibinfo {pages} {182} (\bibinfo {year}
  {2012})}\BibitemShut {NoStop}%
\bibitem [{\citenamefont {Chan}\ and\ \citenamefont
  {Sharma}(2011)}]{chanreview}%
  \BibitemOpen
  \bibfield  {author} {\bibinfo {author} {\bibfnamefont {G.~K.-L.}\
  \bibnamefont {Chan}}\ and\ \bibinfo {author} {\bibfnamefont {S.}~\bibnamefont
  {Sharma}},\ }\href@noop {} {\bibfield  {journal} {\bibinfo  {journal} {Annu.
  Rev. Phys. Chem.}\ }\textbf {\bibinfo {volume} {62}},\ \bibinfo {pages} {465}
  (\bibinfo {year} {2011})}\BibitemShut {NoStop}%
\bibitem [{\citenamefont {Pollet}(2012)}]{Pollet2012}%
  \BibitemOpen
  \bibfield  {author} {\bibinfo {author} {\bibfnamefont {L.}~\bibnamefont
  {Pollet}},\ }\href@noop {} {\bibfield  {journal} {\bibinfo  {journal} {Rep.
  Prog. Phys.}\ }\textbf {\bibinfo {volume} {75}},\ \bibinfo {pages} {094501}
  (\bibinfo {year} {2012})}\BibitemShut {NoStop}%
\bibitem [{\citenamefont {Rodr\'{\i}guez-Guzm\'{a}n}\ \emph
  {et~al.}(2013)\citenamefont {Rodr\'{\i}guez-Guzm\'{a}n}, \citenamefont
  {Jim\'{e}nez-Hoyos}, \citenamefont {Schutski},\ and\ \citenamefont
  {Scuseria}}]{Hubbard-Gustavo_2013}%
  \BibitemOpen
  \bibfield  {author} {\bibinfo {author} {\bibfnamefont {R.}~\bibnamefont
  {Rodr\'{\i}guez-Guzm\'{a}n}}, \bibinfo {author} {\bibfnamefont {C.~A.}\
  \bibnamefont {Jim\'{e}nez-Hoyos}}, \bibinfo {author} {\bibfnamefont
  {R.}~\bibnamefont {Schutski}}, \ and\ \bibinfo {author} {\bibfnamefont
  {G.~E.}\ \bibnamefont {Scuseria}},\ }\href@noop {} {\bibfield  {journal}
  {\bibinfo  {journal} {Phys. Rev. B}\ }\textbf {\bibinfo {volume} {87}},\
  \bibinfo {pages} {235129} (\bibinfo {year} {2013})}\BibitemShut {NoStop}%
\bibitem [{\citenamefont {White}(1992)}]{white}%
  \BibitemOpen
  \bibfield  {author} {\bibinfo {author} {\bibfnamefont {S.~R.}\ \bibnamefont
  {White}},\ }\href@noop {} {\bibfield  {journal} {\bibinfo  {journal} {Phys.
  Rev. Lett.}\ }\textbf {\bibinfo {volume} {69}},\ \bibinfo {pages} {2863}
  (\bibinfo {year} {1992})}\BibitemShut {NoStop}%
\bibitem [{\citenamefont {Schollw\"ock}(2005)}]{scholl05}%
  \BibitemOpen
  \bibfield  {author} {\bibinfo {author} {\bibfnamefont {U.}~\bibnamefont
  {Schollw\"ock}},\ }\href@noop {} {\bibfield  {journal} {\bibinfo  {journal}
  {Rev. Mod. Phys.}\ }\textbf {\bibinfo {volume} {77}},\ \bibinfo {pages} {259}
  (\bibinfo {year} {2005})}\BibitemShut {NoStop}%
\bibitem [{\citenamefont {Legeza}\ \emph {et~al.}(2008)\citenamefont {Legeza},
  \citenamefont {Noack}, \citenamefont {S\'olyom},\ and\ \citenamefont
  {Tincani}}]{ors_springer}%
  \BibitemOpen
  \bibfield  {author} {\bibinfo {author} {\bibfnamefont {O.}~\bibnamefont
  {Legeza}}, \bibinfo {author} {\bibfnamefont {R.~M.}\ \bibnamefont {Noack}},
  \bibinfo {author} {\bibfnamefont {J.}~\bibnamefont {S\'olyom}}, \ and\
  \bibinfo {author} {\bibfnamefont {L.}~\bibnamefont {Tincani}},\ }in\
  \href@noop {} {\emph {\bibinfo {booktitle} {Computational Many-Particle
  Physics}}},\ \bibinfo {series} {Lect. Notes Phys.}, Vol.\ \bibinfo {volume}
  {739},\ \bibinfo {editor} {edited by\ \bibinfo {editor} {\bibfnamefont
  {H.}~\bibnamefont {Fehske}}, \bibinfo {editor} {\bibfnamefont
  {R.}~\bibnamefont {Schneider}}, \ and\ \bibinfo {editor} {\bibfnamefont
  {A.}~\bibnamefont {Wei\ss{}e}}}\ (\bibinfo  {publisher} {Springer},\ \bibinfo
  {address} {Berlin/Heidelerg},\ \bibinfo {year} {2008})\ pp.\ \bibinfo {pages}
  {653--664}\BibitemShut {NoStop}%
\bibitem [{\citenamefont {Marti}\ and\ \citenamefont
  {Reiher}(2010)}]{marti2010b}%
  \BibitemOpen
  \bibfield  {author} {\bibinfo {author} {\bibfnamefont {K.~H.}\ \bibnamefont
  {Marti}}\ and\ \bibinfo {author} {\bibfnamefont {M.}~\bibnamefont {Reiher}},\
  }\href@noop {} {\bibfield  {journal} {\bibinfo  {journal} {Z. Phys. Chem.}\
  }\textbf {\bibinfo {volume} {224}},\ \bibinfo {pages} {583} (\bibinfo {year}
  {2010})}\BibitemShut {NoStop}%
\bibitem [{\citenamefont {Boguslawski}\ \emph
  {et~al.}(2012{\natexlab{a}})\citenamefont {Boguslawski}, \citenamefont
  {Marti}, \citenamefont {Legeza},\ and\ \citenamefont {Reiher}}]{fenoDMRG}%
  \BibitemOpen
  \bibfield  {author} {\bibinfo {author} {\bibfnamefont {K.}~\bibnamefont
  {Boguslawski}}, \bibinfo {author} {\bibfnamefont {K.~H.}\ \bibnamefont
  {Marti}}, \bibinfo {author} {\bibfnamefont {O.}~\bibnamefont {Legeza}}, \
  and\ \bibinfo {author} {\bibfnamefont {M.}~\bibnamefont {Reiher}},\
  }\href@noop {} {\bibfield  {journal} {\bibinfo  {journal} {J. Chem. Theory
  Comput.}\ }\textbf {\bibinfo {volume} {8}},\ \bibinfo {pages} {1970}
  (\bibinfo {year} {2012}{\natexlab{a}})}\BibitemShut {NoStop}%
\bibitem [{\citenamefont {Boguslawski}\ \emph
  {et~al.}(2012{\natexlab{b}})\citenamefont {Boguslawski}, \citenamefont
  {Tecmer}, \citenamefont {Legeza},\ and\ \citenamefont
  {Reiher}}]{entanglement_letter}%
  \BibitemOpen
  \bibfield  {author} {\bibinfo {author} {\bibfnamefont {K.}~\bibnamefont
  {Boguslawski}}, \bibinfo {author} {\bibfnamefont {P.}~\bibnamefont {Tecmer}},
  \bibinfo {author} {\bibfnamefont {O.}~\bibnamefont {Legeza}}, \ and\ \bibinfo
  {author} {\bibfnamefont {M.}~\bibnamefont {Reiher}},\ }\href@noop {}
  {\bibfield  {journal} {\bibinfo  {journal} {J. Phys. Chem. Lett.}\ }\textbf
  {\bibinfo {volume} {3}},\ \bibinfo {pages} {3129} (\bibinfo {year}
  {2012}{\natexlab{b}})}\BibitemShut {NoStop}%
\bibitem [{\citenamefont {Wouters}\ \emph {et~al.}(2012)\citenamefont
  {Wouters}, \citenamefont {Limacher}, \citenamefont {{Van Neck}},\ and\
  \citenamefont {Ayers}}]{DMRG_polarizability}%
  \BibitemOpen
  \bibfield  {author} {\bibinfo {author} {\bibfnamefont {S.}~\bibnamefont
  {Wouters}}, \bibinfo {author} {\bibfnamefont {P.~A.}\ \bibnamefont
  {Limacher}}, \bibinfo {author} {\bibfnamefont {D.}~\bibnamefont {{Van
  Neck}}}, \ and\ \bibinfo {author} {\bibfnamefont {P.~W.}\ \bibnamefont
  {Ayers}},\ }\href@noop {} {\bibfield  {journal} {\bibinfo  {journal} {J.
  Chem. Phys.}\ }\textbf {\bibinfo {volume} {136}},\ \bibinfo {pages} {134110}
  (\bibinfo {year} {2012})}\BibitemShut {NoStop}%
\bibitem [{\citenamefont {Kurashige}\ \emph {et~al.}(2013)\citenamefont
  {Kurashige}, \citenamefont {Chan},\ and\ \citenamefont
  {Yanai}}]{kurashige2013}%
  \BibitemOpen
  \bibfield  {author} {\bibinfo {author} {\bibfnamefont {Y.}~\bibnamefont
  {Kurashige}}, \bibinfo {author} {\bibfnamefont {G.~K.-L.}\ \bibnamefont
  {Chan}}, \ and\ \bibinfo {author} {\bibfnamefont {T.}~\bibnamefont {Yanai}},\
  }\href@noop {} {\bibfield  {journal} {\bibinfo  {journal} {Nature Chem.}\
  }\textbf {\bibinfo {volume} {5}},\ \bibinfo {pages} {660} (\bibinfo {year}
  {2013})}\BibitemShut {NoStop}%
\bibitem [{\citenamefont {Boguslawski}\ \emph {et~al.}(2013)\citenamefont
  {Boguslawski}, \citenamefont {Tecmer}, \citenamefont {Barcza}, \citenamefont
  {Legeza},\ and\ \citenamefont {Reiher}}]{entanglement_bonding_2013}%
  \BibitemOpen
  \bibfield  {author} {\bibinfo {author} {\bibfnamefont {K.}~\bibnamefont
  {Boguslawski}}, \bibinfo {author} {\bibfnamefont {P.}~\bibnamefont {Tecmer}},
  \bibinfo {author} {\bibfnamefont {G.}~\bibnamefont {Barcza}}, \bibinfo
  {author} {\bibfnamefont {O.}~\bibnamefont {Legeza}}, \ and\ \bibinfo {author}
  {\bibfnamefont {M.}~\bibnamefont {Reiher}},\ }\href@noop {} {\bibfield
  {journal} {\bibinfo  {journal} {J. Chem. Theory Comput.}\ }\textbf {\bibinfo
  {volume} {9}},\ \bibinfo {pages} {2959} (\bibinfo {year} {2013})}\BibitemShut
  {NoStop}%
\bibitem [{\citenamefont {Tecmer}\ \emph {et~al.}(2014)\citenamefont {Tecmer},
  \citenamefont {Boguslawski}, \citenamefont {Legeza},\ and\ \citenamefont
  {Reiher}}]{CUO_DMRG}%
  \BibitemOpen
  \bibfield  {author} {\bibinfo {author} {\bibfnamefont {P.}~\bibnamefont
  {Tecmer}}, \bibinfo {author} {\bibfnamefont {K.}~\bibnamefont {Boguslawski}},
  \bibinfo {author} {\bibfnamefont {O.}~\bibnamefont {Legeza}}, \ and\ \bibinfo
  {author} {\bibfnamefont {M.}~\bibnamefont {Reiher}},\ }\href@noop {}
  {\bibfield  {journal} {\bibinfo  {journal} {Phys. Chem. Chem. Phys}\ }\textbf
  {\bibinfo {volume} {16}},\ \bibinfo {pages} {719} (\bibinfo {year}
  {2014})}\BibitemShut {NoStop}%
\bibitem [{\citenamefont {Wouters}\ \emph {et~al.}(2014)\citenamefont
  {Wouters}, \citenamefont {Poelmans}, \citenamefont {Ayers},\ and\
  \citenamefont {{Van Neck}}}]{CheMPS2}%
  \BibitemOpen
  \bibfield  {author} {\bibinfo {author} {\bibfnamefont {S.}~\bibnamefont
  {Wouters}}, \bibinfo {author} {\bibfnamefont {W.}~\bibnamefont {Poelmans}},
  \bibinfo {author} {\bibfnamefont {P.~W.}\ \bibnamefont {Ayers}}, \ and\
  \bibinfo {author} {\bibfnamefont {D.}~\bibnamefont {{Van Neck}}},\
  }\href@noop {} {\bibfield  {journal} {\bibinfo  {journal} {Comput. Phys.
  Comm.}\ }\textbf {\bibinfo {volume} {XX}},\ \bibinfo {pages}
  {DOI:10.1016/j.cpc.2014.01.019} (\bibinfo {year} {2014})}\BibitemShut
  {NoStop}%
\bibitem [{\citenamefont {Chung}\ and\ \citenamefont
  {Peschel}(2000)}]{Chung2000}%
  \BibitemOpen
  \bibfield  {author} {\bibinfo {author} {\bibfnamefont {M.~C.}\ \bibnamefont
  {Chung}}\ and\ \bibinfo {author} {\bibfnamefont {I.}~\bibnamefont
  {Peschel}},\ }\href@noop {} {\bibfield  {journal} {\bibinfo  {journal} {Phys.
  Rev. B}\ }\textbf {\bibinfo {volume} {62}},\ \bibinfo {pages} {4191}
  (\bibinfo {year} {2000})}\BibitemShut {NoStop}%
\bibitem [{\citenamefont {Murg}\ \emph {et~al.}(2010)\citenamefont {Murg},
  \citenamefont {Verstraete}, \citenamefont {Legeza},\ and\ \citenamefont
  {Noack}}]{TTN_Ors}%
  \BibitemOpen
  \bibfield  {author} {\bibinfo {author} {\bibfnamefont {V.}~\bibnamefont
  {Murg}}, \bibinfo {author} {\bibfnamefont {F.}~\bibnamefont {Verstraete}},
  \bibinfo {author} {\bibfnamefont {O.}~\bibnamefont {Legeza}}, \ and\ \bibinfo
  {author} {\bibfnamefont {R.~M.}\ \bibnamefont {Noack}},\ }\href@noop {}
  {\bibfield  {journal} {\bibinfo  {journal} {Phys. Rev. B}\ }\textbf {\bibinfo
  {volume} {82}},\ \bibinfo {pages} {205105} (\bibinfo {year}
  {2010})}\BibitemShut {NoStop}%
\bibitem [{\citenamefont {Hurley}\ \emph {et~al.}(1953)\citenamefont {Hurley},
  \citenamefont {Lennard-Jones},\ and\ \citenamefont {Pople}}]{Hurley_1953}%
  \BibitemOpen
  \bibfield  {author} {\bibinfo {author} {\bibfnamefont {A.~C.}\ \bibnamefont
  {Hurley}}, \bibinfo {author} {\bibfnamefont {J.}~\bibnamefont
  {Lennard-Jones}}, \ and\ \bibinfo {author} {\bibfnamefont {J.~A.}\
  \bibnamefont {Pople}},\ }\href@noop {} {\bibfield  {journal} {\bibinfo
  {journal} {Proc. R. Soc. Lond. A}\ }\textbf {\bibinfo {volume} {220}},\
  \bibinfo {pages} {446} (\bibinfo {year} {1953})}\BibitemShut {NoStop}%
\bibitem [{\citenamefont {Coleman}(1965)}]{Coleman_1965}%
  \BibitemOpen
  \bibfield  {author} {\bibinfo {author} {\bibfnamefont {A.~J.}\ \bibnamefont
  {Coleman}},\ }\href@noop {} {\bibfield  {journal} {\bibinfo  {journal} {J.
  Math. Phys.}\ }\textbf {\bibinfo {volume} {6}},\ \bibinfo {pages} {1425}
  (\bibinfo {year} {1965})}\BibitemShut {NoStop}%
\bibitem [{\citenamefont {Silver}(1969)}]{Silver_1969}%
  \BibitemOpen
  \bibfield  {author} {\bibinfo {author} {\bibfnamefont {D.~M.}\ \bibnamefont
  {Silver}},\ }\href@noop {} {\bibfield  {journal} {\bibinfo  {journal} {J.
  Chem. Phys.}\ }\textbf {\bibinfo {volume} {50}},\ \bibinfo {pages} {5108}
  (\bibinfo {year} {1969})}\BibitemShut {NoStop}%
\bibitem [{\citenamefont {Ortiz}\ \emph {et~al.}(1981)\citenamefont {Ortiz},
  \citenamefont {Weiner},\ and\ \citenamefont {Ohrn}}]{Ortiz_1981}%
  \BibitemOpen
  \bibfield  {author} {\bibinfo {author} {\bibfnamefont {J.~V.}\ \bibnamefont
  {Ortiz}}, \bibinfo {author} {\bibfnamefont {B.}~\bibnamefont {Weiner}}, \
  and\ \bibinfo {author} {\bibfnamefont {Y.}~\bibnamefont {Ohrn}},\ }\href@noop
  {} {\bibfield  {journal} {\bibinfo  {journal} {Int. J. Quantum Chem.}\
  }\textbf {\bibinfo {volume} {S15}},\ \bibinfo {pages} {113} (\bibinfo {year}
  {1981})}\BibitemShut {NoStop}%
\bibitem [{\citenamefont {Surjan}(1999)}]{Surjan_1999}%
  \BibitemOpen
  \bibfield  {author} {\bibinfo {author} {\bibfnamefont {P.~R.}\ \bibnamefont
  {Surjan}},\ }in\ \href@noop {} {\emph {\bibinfo {booktitle} {Correlation and
  Localization}}}\ (\bibinfo  {publisher} {Springer},\ \bibinfo {year} {1999})\
  pp.\ \bibinfo {pages} {63--88}\BibitemShut {NoStop}%
\bibitem [{\citenamefont {Kutzelnigg}(2012)}]{Kutzelnigg2012}%
  \BibitemOpen
  \bibfield  {author} {\bibinfo {author} {\bibfnamefont {W.}~\bibnamefont
  {Kutzelnigg}},\ }\href@noop {} {\bibfield  {journal} {\bibinfo  {journal}
  {Chem. Phys.}\ }\textbf {\bibinfo {volume} {401}},\ \bibinfo {pages} {119}
  (\bibinfo {year} {2012})}\BibitemShut {NoStop}%
\bibitem [{\citenamefont {Surj\'{a}n}\ \emph {et~al.}(2012)\citenamefont
  {Surj\'{a}n}, \citenamefont {Szabados}, \citenamefont {Jeszenszki},\ and\
  \citenamefont {Zoboki}}]{Surjan_2012}%
  \BibitemOpen
  \bibfield  {author} {\bibinfo {author} {\bibfnamefont {P.~R.}\ \bibnamefont
  {Surj\'{a}n}}, \bibinfo {author} {\bibfnamefont {A.}~\bibnamefont
  {Szabados}}, \bibinfo {author} {\bibfnamefont {P.}~\bibnamefont
  {Jeszenszki}}, \ and\ \bibinfo {author} {\bibfnamefont {T.}~\bibnamefont
  {Zoboki}},\ }\href@noop {} {\bibfield  {journal} {\bibinfo  {journal} {J.
  Math. Chem.}\ }\textbf {\bibinfo {volume} {50}},\ \bibinfo {pages} {534}
  (\bibinfo {year} {2012})}\BibitemShut {NoStop}%
\bibitem [{\citenamefont {Ellis}\ \emph {et~al.}(2013)\citenamefont {Ellis},
  \citenamefont {Martin},\ and\ \citenamefont {Scuseria}}]{Ellis_2013}%
  \BibitemOpen
  \bibfield  {author} {\bibinfo {author} {\bibfnamefont {J.~K.}\ \bibnamefont
  {Ellis}}, \bibinfo {author} {\bibfnamefont {R.~L.}\ \bibnamefont {Martin}}, \
  and\ \bibinfo {author} {\bibfnamefont {G.~E.}\ \bibnamefont {Scuseria}},\
  }\href@noop {} {\bibfield  {journal} {\bibinfo  {journal} {J. Chem. Theory
  Comput.}\ }\textbf {\bibinfo {volume} {9}},\ \bibinfo {pages} {2857}
  (\bibinfo {year} {2013})}\BibitemShut {NoStop}%
\bibitem [{\citenamefont {Limacher}\ \emph {et~al.}(2013)\citenamefont
  {Limacher}, \citenamefont {Ayers}, \citenamefont {Johnson}, \citenamefont
  {{De Baerdemacker}}, \citenamefont {{Van Neck}},\ and\ \citenamefont
  {Bultinck}}]{Limacher_2013}%
  \BibitemOpen
  \bibfield  {author} {\bibinfo {author} {\bibfnamefont {P.~A.}\ \bibnamefont
  {Limacher}}, \bibinfo {author} {\bibfnamefont {P.~W.}\ \bibnamefont {Ayers}},
  \bibinfo {author} {\bibfnamefont {P.~A.}\ \bibnamefont {Johnson}}, \bibinfo
  {author} {\bibfnamefont {S.}~\bibnamefont {{De Baerdemacker}}}, \bibinfo
  {author} {\bibfnamefont {D.}~\bibnamefont {{Van Neck}}}, \ and\ \bibinfo
  {author} {\bibfnamefont {P.}~\bibnamefont {Bultinck}},\ }\href@noop {}
  {\bibfield  {journal} {\bibinfo  {journal} {J. Chem. Theory Comput.}\
  }\textbf {\bibinfo {volume} {9}},\ \bibinfo {pages} {1394} (\bibinfo {year}
  {2013})}\BibitemShut {NoStop}%
\bibitem [{\citenamefont {Limacher}\ \emph
  {et~al.}(2014{\natexlab{a}})\citenamefont {Limacher}, \citenamefont {Ayers},
  \citenamefont {Johnson}, \citenamefont {{De Baerdemacker}}, \citenamefont
  {{Van Neck}},\ and\ \citenamefont {Bultinck}}]{Piotrus_PT2}%
  \BibitemOpen
  \bibfield  {author} {\bibinfo {author} {\bibfnamefont {P.}~\bibnamefont
  {Limacher}}, \bibinfo {author} {\bibfnamefont {P.}~\bibnamefont {Ayers}},
  \bibinfo {author} {\bibfnamefont {P.}~\bibnamefont {Johnson}}, \bibinfo
  {author} {\bibfnamefont {S.}~\bibnamefont {{De Baerdemacker}}}, \bibinfo
  {author} {\bibfnamefont {D.}~\bibnamefont {{Van Neck}}}, \ and\ \bibinfo
  {author} {\bibfnamefont {P.}~\bibnamefont {Bultinck}},\ }\href@noop {}
  {\bibfield  {journal} {\bibinfo  {journal} {Phys. Chem. Chem. Phys}\ }\textbf
  {\bibinfo {volume} {16}},\ \bibinfo {pages} {5061} (\bibinfo {year}
  {2014}{\natexlab{a}})}\BibitemShut {NoStop}%
\bibitem [{\citenamefont {Limacher}\ \emph
  {et~al.}(2014{\natexlab{b}})\citenamefont {Limacher}, \citenamefont {Kim},
  \citenamefont {Ayers}, \citenamefont {Johnson}, \citenamefont {{De
  Baerdemacker}}, \citenamefont {Van~Neck},\ and\ \citenamefont
  {Bultinck}}]{Piotrus_Mol-Phys}%
  \BibitemOpen
  \bibfield  {author} {\bibinfo {author} {\bibfnamefont {P.~A.}\ \bibnamefont
  {Limacher}}, \bibinfo {author} {\bibfnamefont {T.~D.}\ \bibnamefont {Kim}},
  \bibinfo {author} {\bibfnamefont {P.~W.}\ \bibnamefont {Ayers}}, \bibinfo
  {author} {\bibfnamefont {P.~A.}\ \bibnamefont {Johnson}}, \bibinfo {author}
  {\bibfnamefont {S.}~\bibnamefont {{De Baerdemacker}}}, \bibinfo {author}
  {\bibfnamefont {D.}~\bibnamefont {Van~Neck}}, \ and\ \bibinfo {author}
  {\bibfnamefont {P.}~\bibnamefont {Bultinck}},\ }\href@noop {} {\bibfield
  {journal} {\bibinfo  {journal} {Mol. Phys.}\ ,\ \bibinfo {pages} {853}}
  (\bibinfo {year} {2014}{\natexlab{b}})}\BibitemShut {NoStop}%
\bibitem [{\citenamefont {Henderson}\ \emph {et~al.}(2014)\citenamefont
  {Henderson}, \citenamefont {Dukelsky}, \citenamefont {Scuseria},
  \citenamefont {Signoracci},\ and\ \citenamefont {Duguet}}]{p-CCD}%
  \BibitemOpen
  \bibfield  {author} {\bibinfo {author} {\bibfnamefont {T.~M.}\ \bibnamefont
  {Henderson}}, \bibinfo {author} {\bibfnamefont {J.}~\bibnamefont {Dukelsky}},
  \bibinfo {author} {\bibfnamefont {G.~E.}\ \bibnamefont {Scuseria}}, \bibinfo
  {author} {\bibfnamefont {A.}~\bibnamefont {Signoracci}}, \ and\ \bibinfo
  {author} {\bibfnamefont {T.}~\bibnamefont {Duguet}},\ }\href@noop {}
  {\bibfield  {journal} {\bibinfo  {journal} {arXiv preprint arXiv:1403.6818}\
  } (\bibinfo {year} {2014})}\BibitemShut {NoStop}%
\bibitem [{\citenamefont {Helgaker}\ \emph {et~al.}(2000)\citenamefont
  {Helgaker}, \citenamefont {J{\o}rgensen},\ and\ \citenamefont
  {Olsen}}]{Helgaker_book}%
  \BibitemOpen
  \bibfield  {author} {\bibinfo {author} {\bibfnamefont {T.}~\bibnamefont
  {Helgaker}}, \bibinfo {author} {\bibfnamefont {P.}~\bibnamefont
  {J{\o}rgensen}}, \ and\ \bibinfo {author} {\bibfnamefont {J.}~\bibnamefont
  {Olsen}},\ }\href@noop {} {\emph {\bibinfo {title} {Molecular
  Electronic-Structure Theory}}}\ (\bibinfo  {publisher} {Wiley},\ \bibinfo
  {year} {2000})\BibitemShut {NoStop}%
\bibitem [{\citenamefont {Scuseria}\ and\ \citenamefont {{Schaefer
  III}}(1987)}]{Scuseria1987}%
  \BibitemOpen
  \bibfield  {author} {\bibinfo {author} {\bibfnamefont {G.~E.}\ \bibnamefont
  {Scuseria}}\ and\ \bibinfo {author} {\bibfnamefont {H.~F.}\ \bibnamefont
  {{Schaefer III}}},\ }\href@noop {} {\bibfield  {journal} {\bibinfo  {journal}
  {Chem. Phys. Lett.}\ }\textbf {\bibinfo {volume} {142}},\ \bibinfo {pages}
  {354} (\bibinfo {year} {1987})}\BibitemShut {NoStop}%
\bibitem [{\citenamefont {Bozkaya}\ \emph {et~al.}(2011)\citenamefont
  {Bozkaya}, \citenamefont {Turney}, \citenamefont {Yamaguchi}, \citenamefont
  {Schaefer},\ and\ \citenamefont {Sherrill}}]{Bozkaya_2011}%
  \BibitemOpen
  \bibfield  {author} {\bibinfo {author} {\bibfnamefont {U.}~\bibnamefont
  {Bozkaya}}, \bibinfo {author} {\bibfnamefont {J.~M.}\ \bibnamefont {Turney}},
  \bibinfo {author} {\bibfnamefont {Y.}~\bibnamefont {Yamaguchi}}, \bibinfo
  {author} {\bibfnamefont {H.~F.}\ \bibnamefont {Schaefer}}, \ and\ \bibinfo
  {author} {\bibfnamefont {C.~D.}\ \bibnamefont {Sherrill}},\ }\href@noop {}
  {\bibfield  {journal} {\bibinfo  {journal} {J. Chem. Phys.}\ }\textbf
  {\bibinfo {volume} {135}},\ \bibinfo {pages} {104103} (\bibinfo {year}
  {2011})}\BibitemShut {NoStop}%
\bibitem [{HOR()}]{HORTON13}%
  \BibitemOpen
  \href@noop {} {}\bibinfo {note} {Horton 1.2.0 2013, written by T.
  Verstraelen, S. Vandenbrande, M. Chan, F. H. Zadeh, C. Gonzalez, K.
  Boguslawski, P. Tecmer, P. A. Limacher, A. Malek (see {\tt
  http://theochem.github.com/horton/})}\BibitemShut {NoStop}%
\bibitem [{\citenamefont {Lieb}\ and\ \citenamefont
  {Wu}(1968{\natexlab{a}})}]{Lieb-Wu}%
  \BibitemOpen
  \bibfield  {author} {\bibinfo {author} {\bibfnamefont {E.~H.}\ \bibnamefont
  {Lieb}}\ and\ \bibinfo {author} {\bibfnamefont {F.~Y.}\ \bibnamefont {Wu}},\
  }\href@noop {} {\bibfield  {journal} {\bibinfo  {journal} {Phys. Rev. Lett.}\
  }\textbf {\bibinfo {volume} {20}},\ \bibinfo {pages} {1445} (\bibinfo {year}
  {1968}{\natexlab{a}})}\BibitemShut {NoStop}%
\bibitem [{\citenamefont {Neuscamman}(2012)}]{Neuscamman_2012}%
  \BibitemOpen
  \bibfield  {author} {\bibinfo {author} {\bibfnamefont {E.}~\bibnamefont
  {Neuscamman}},\ }\href@noop {} {\bibfield  {journal} {\bibinfo  {journal}
  {Phys. Rev. Lett.}\ }\textbf {\bibinfo {volume} {109}},\ \bibinfo {pages}
  {203001} (\bibinfo {year} {2012})}\BibitemShut {NoStop}%
\bibitem [{\citenamefont {Bytautas}\ \emph {et~al.}(2011)\citenamefont
  {Bytautas}, \citenamefont {Henderson}, \citenamefont {Jim\'{e}nez-Hoyos},
  \citenamefont {Ellis},\ and\ \citenamefont {Scuseria}}]{Gus_seniority}%
  \BibitemOpen
  \bibfield  {author} {\bibinfo {author} {\bibfnamefont {L.}~\bibnamefont
  {Bytautas}}, \bibinfo {author} {\bibfnamefont {T.~M.}\ \bibnamefont
  {Henderson}}, \bibinfo {author} {\bibfnamefont {C.~A.}\ \bibnamefont
  {Jim\'{e}nez-Hoyos}}, \bibinfo {author} {\bibfnamefont {J.~K.}\ \bibnamefont
  {Ellis}}, \ and\ \bibinfo {author} {\bibfnamefont {G.~E.}\ \bibnamefont
  {Scuseria}},\ }\href@noop {} {\bibfield  {journal} {\bibinfo  {journal} {J.
  Chem. Phys.}\ }\textbf {\bibinfo {volume} {135}},\ \bibinfo {pages} {044119}
  (\bibinfo {year} {2011})}\BibitemShut {NoStop}%
\bibitem [{\citenamefont {Bethe}(1931)}]{Bethe_ansatz}%
  \BibitemOpen
  \bibfield  {author} {\bibinfo {author} {\bibfnamefont {H.}~\bibnamefont
  {Bethe}},\ }\href@noop {} {\bibfield  {journal} {\bibinfo  {journal} {Z.
  Phys.}\ }\textbf {\bibinfo {volume} {71}},\ \bibinfo {pages} {205} (\bibinfo
  {year} {1931})}\BibitemShut {NoStop}%
\bibitem [{\citenamefont {Lieb}\ and\ \citenamefont
  {Wu}(1968{\natexlab{b}})}]{Lieb_1968}%
  \BibitemOpen
  \bibfield  {author} {\bibinfo {author} {\bibfnamefont {E.~H.}\ \bibnamefont
  {Lieb}}\ and\ \bibinfo {author} {\bibfnamefont {F.~Y.}\ \bibnamefont {Wu}},\
  }\href@noop {} {\bibfield  {journal} {\bibinfo  {journal} {Phys. Rev. Lett.}\
  }\textbf {\bibinfo {volume} {20}},\ \bibinfo {pages} {1445} (\bibinfo {year}
  {1968}{\natexlab{b}})}\BibitemShut {NoStop}%
\bibitem [{\citenamefont {Gu}\ \emph {et~al.}(2004)\citenamefont {Gu},
  \citenamefont {Deng}, \citenamefont {Li},\ and\ \citenamefont
  {Lin}}]{Gu2004}%
  \BibitemOpen
  \bibfield  {author} {\bibinfo {author} {\bibfnamefont {S.-J.}\ \bibnamefont
  {Gu}}, \bibinfo {author} {\bibfnamefont {S.-S.}\ \bibnamefont {Deng}},
  \bibinfo {author} {\bibfnamefont {Y.-Q.}\ \bibnamefont {Li}}, \ and\ \bibinfo
  {author} {\bibfnamefont {H.-Q.}\ \bibnamefont {Lin}},\ }\href@noop {}
  {\bibfield  {journal} {\bibinfo  {journal} {Phys. Rev. Lett.}\ }\textbf
  {\bibinfo {volume} {93}},\ \bibinfo {pages} {086402} (\bibinfo {year}
  {2004})}\BibitemShut {NoStop}%
\bibitem [{\citenamefont {Hachmann}\ \emph {et~al.}(2006)\citenamefont
  {Hachmann}, \citenamefont {Cardoen},\ and\ \citenamefont
  {Chan}}]{Hachmann_H50}%
  \BibitemOpen
  \bibfield  {author} {\bibinfo {author} {\bibfnamefont {J.}~\bibnamefont
  {Hachmann}}, \bibinfo {author} {\bibfnamefont {W.}~\bibnamefont {Cardoen}}, \
  and\ \bibinfo {author} {\bibfnamefont {G.~K.-L.}\ \bibnamefont {Chan}},\
  }\href@noop {} {\bibfield  {journal} {\bibinfo  {journal} {J. Chem. Phys.}\
  }\textbf {\bibinfo {volume} {125}},\ \bibinfo {pages} {144101} (\bibinfo
  {year} {2006})}\BibitemShut {NoStop}%
\bibitem [{\citenamefont {Tsuchimochi}\ and\ \citenamefont
  {Scuseria}(2009)}]{Gustavo_H50}%
  \BibitemOpen
  \bibfield  {author} {\bibinfo {author} {\bibfnamefont {T.}~\bibnamefont
  {Tsuchimochi}}\ and\ \bibinfo {author} {\bibfnamefont {G.~E.}\ \bibnamefont
  {Scuseria}},\ }\href@noop {} {\bibfield  {journal} {\bibinfo  {journal} {J.
  Chem. Phys.}\ }\textbf {\bibinfo {volume} {131}},\ \bibinfo {pages} {121102}
  (\bibinfo {year} {2009})}\BibitemShut {NoStop}%
\bibitem [{\citenamefont {Stella}\ \emph {et~al.}(2011)\citenamefont {Stella},
  \citenamefont {Attaccalite}, \citenamefont {Sorella},\ and\ \citenamefont
  {Rubio}}]{Stella_H50}%
  \BibitemOpen
  \bibfield  {author} {\bibinfo {author} {\bibfnamefont {L.}~\bibnamefont
  {Stella}}, \bibinfo {author} {\bibfnamefont {C.}~\bibnamefont {Attaccalite}},
  \bibinfo {author} {\bibfnamefont {S.}~\bibnamefont {Sorella}}, \ and\
  \bibinfo {author} {\bibfnamefont {A.}~\bibnamefont {Rubio}},\ }\href@noop {}
  {\bibfield  {journal} {\bibinfo  {journal} {Phys. Rev. B}\ }\textbf {\bibinfo
  {volume} {84}},\ \bibinfo {pages} {245117} (\bibinfo {year}
  {2011})}\BibitemShut {NoStop}%
\bibitem [{\citenamefont {Lin}\ \emph {et~al.}(2011)\citenamefont {Lin},
  \citenamefont {A.~Marianetti}, \citenamefont {Millis},\ and\ \citenamefont
  {Reichman}}]{DMFT_H50}%
  \BibitemOpen
  \bibfield  {author} {\bibinfo {author} {\bibfnamefont {N.}~\bibnamefont
  {Lin}}, \bibinfo {author} {\bibfnamefont {C.}~\bibnamefont {A.~Marianetti}},
  \bibinfo {author} {\bibfnamefont {A.~J.}\ \bibnamefont {Millis}}, \ and\
  \bibinfo {author} {\bibfnamefont {D.~R.}\ \bibnamefont {Reichman}},\
  }\href@noop {} {\bibfield  {journal} {\bibinfo  {journal} {Phys. Rev. Lett.}\
  }\textbf {\bibinfo {volume} {106}},\ \bibinfo {pages} {096402} (\bibinfo
  {year} {2011})}\BibitemShut {NoStop}%
\bibitem [{\citenamefont {Hehre}\ \emph {et~al.}(1969)\citenamefont {Hehre},
  \citenamefont {Stewart},\ and\ \citenamefont {Pople}}]{STO-6G-H-Ne}%
  \BibitemOpen
  \bibfield  {author} {\bibinfo {author} {\bibfnamefont {W.~J.}\ \bibnamefont
  {Hehre}}, \bibinfo {author} {\bibfnamefont {R.~F.}\ \bibnamefont {Stewart}},
  \ and\ \bibinfo {author} {\bibfnamefont {J.~A.}\ \bibnamefont {Pople}},\
  }\href@noop {} {\bibfield  {journal} {\bibinfo  {journal} {J. Chem. Phys.}\
  }\textbf {\bibinfo {volume} {51}},\ \bibinfo {pages} {2657} (\bibinfo {year}
  {1969})}\BibitemShut {NoStop}%
\bibitem [{\citenamefont {Perdew}\ \emph {et~al.}(1996)\citenamefont {Perdew},
  \citenamefont {Burke},\ and\ \citenamefont {Ernzerhof}}]{PBEx}%
  \BibitemOpen
  \bibfield  {author} {\bibinfo {author} {\bibfnamefont {J.~P.}\ \bibnamefont
  {Perdew}}, \bibinfo {author} {\bibfnamefont {K.}~\bibnamefont {Burke}}, \
  and\ \bibinfo {author} {\bibfnamefont {M.}~\bibnamefont {Ernzerhof}},\
  }\href@noop {} {\bibfield  {journal} {\bibinfo  {journal} {Phys. Rev. Lett.}\
  }\textbf {\bibinfo {volume} {77}},\ \bibinfo {pages} {3865} (\bibinfo {year}
  {1996})}\BibitemShut {NoStop}%
\end{thebibliography}%
\end{document}